\documentclass[12pt,a4paper]{article}
\usepackage{a4wide}
\usepackage{latexsym}
\usepackage{epsf}
\usepackage{amssymb}
\topmargin
-1cm

\def\a{\alpha}
\def\b{\beta}

\def\d{\delta}
\def\m{\mu}
\def\n{\nu}

\def\L{\Lambda}
\def\s{\sigma}

\def\be{\begin{equation}}
\def\ee{\end{equation}}
\def\bea{\begin{eqnarray}}
\def\eea{\end{eqnarray}}

\begin{document}
\begin{flushright}
IFT-UAM/CSIC-01-06\\
hep-th/0102130\\
\end{flushright}

\vspace{1cm}

\begin{center}

{\bf\Large  Dilatonic Wilson Loops}

\vspace{.5cm}

{\bf Juan Jos\'e Manjar\'{\i}n}
\footnote{E-mail: {\tt juanjose.manjarin@uam.es}} \\
\vspace{.3cm}

\vskip 0.4cm
 {\it   
Instituto de F\'{\i}sica Te\'orica, C-XVI,
  Universidad Aut\'onoma de Madrid \\
  E-28049-Madrid, Spain}\footnote{Unidad de Investigaci\'on Asociada
  al Centro de F\'{\i}sica Miguel Catal\'an (C.S.I.C.)}

\vskip 0.2cm

and
\vskip 0.4cm
{\it Departamento de F\'{\i}sica Te\'orica, C-XI,
  Universidad Aut\'onoma de Madrid \\
  E-28049-Madrid, Spain }

\vskip 0.2cm

\vskip 1cm


{\bf Abstract}
\end{center}
In this letter we study the dilatonic corrections to the static gauge potential between heavy sources. These corrections come from the solutions to the the lowest order beta equations. In the energetically favoured branch, the potential obtained is characterised by having a linear confining term, an $L$ independent term and another $1/L$ piece. This is indicative of a L\"uscher-type behaviour in the strong-coupling regime of the dual gauge theory. On the other hand, we also explore the singularity as a point where the theory becomes free.

\newpage
\section{Introduction}

In the recent years there have been a renewal interest in the study of the conexion between strings and gauge theories, mainly since the conjecture relating type IIB string theory on AdS$_5\times$S$^5$, {\cite{conjet}}, was stated. This conjecture is related to the dynamics of a system of $N$ D3-branes in the decouplig limit $\a'\rightarrow 0$ and can be seen as a strong-weak duality, where the strong t'Hooft coupling in the large $N$ limit is in correspondence with weakly coupled type IIB string theory.

However, despite its great insight, the Maldacena's conjecture is related to a non-realistic gauge theory, namely, $U(N)$ $N=4$ super Yang-Mills, and so a further exploration is needed. The analysis can be extended to obtain a nonsupersymmetric gauge theory by discussing the finite temperature behaviour of the $N=4$ gauge theory \cite{ft}. In this case the time direction of the AdS background is compactified on a circle of radius $\propto 1/T$, so that supersymmetry is broken by the boundary conditions on the circle.

Another way to get a non-supersymmetric Yang-Mills theory is through the notion of confining strings {\cite{confstr}}. Phenomenologically, this idea relies on the interpretation of the color flux tubes responsible for quark confinement as strings ending on quark-antiquark pairs.

One way to achieve a string representation of this configuration is based on the holographic renormalization group approach \cite{rg}. The procedure is to model out the renormalization group equations of the quantum field theory from the data coming from string theory, that is, the equations for the running of the coupling constants of the 4d non-abelian gauge theory are related to the Einstein equations of a 5d system of gravity plus matter. In this way, certain background fields of a closed string theory are associated to the coupling constants of the field theory. Thus geometry dictates the properties of the gauge theory.

This five dimensional theory is IR divergent due to the contributions of the boundary $\rho=\infty$, which corresponds to UV divergences of the four dimensional gauge theory. This is the origin of the IR/UV connection. In order to put these divergences under control we introduce a regulator at a scale $\rho=\Lambda$ and demand that the physical observables be independent on the particular choice of $\Lambda$. This implements the Callan-Symanzik equations of the corresponding QFT.

In \cite{ag} was shown that the dynamics underlying the Callan-Symanzik equations is just the conditions for the conformal invariance of the two-dimensional string sigma model and the fact that open string one loop divergences just happen to be represented by insertions of closed string vertex operators.

The conditions for conformal invariance are just given by setting the beta equations of the string to zero. Assuming a vanishing Kalb-Ramond field, the beta equations for a closed bosonic string theory read as

\bea
\label{beta1}
R_{\mu\nu}-\nabla_{\mu}\nabla_{\nu}\Phi=0,\\
\label{beta2}
\nabla^2\Phi+(\nabla\Phi)^2=\frac{26-D}{3\alpha'},
\eea

\noindent where $D$ is the dimension of the space-time. 

In \cite{agh} it has been shown that in the context of non-critical strings there is a solution to these equations, to first order in $\alpha'$, for generic values of D compatible with $ISO(1,D-1)$ Poincar\'e invariance. That is 

\be
\label{metlor}
ds^2={\mbox{tanh}}\left(\sqrt{\frac{21}{6\alpha'}}r\right)^{2/\sqrt{d}}\eta_{\mu\nu}dx^{\mu}dx^{\nu}+dr^2,
\ee
\be
\label{dillor}
\Phi=-\log\left[{\mbox{tanh}}\left(\sqrt{\frac{21}{6\alpha'}}r\right)^{(d-2)/2}\frac{1}{{\mbox{cosh}}^2\left(\sqrt{\frac{21}{6\alpha'}}r\right)}\right],
\ee

\noindent where $r$ is interpreted as the holographic coordinate or the Liouville direction, depending on the context.

One of the main features of this solution is its validity for any number of space-time dimensions, which allows us to work directly with a 5d string background with no need to make further assumptions on the behaviour of the towers of states coming form the excitations of the string on the 21 extra dimensions. In this way, from now on we will work in a $D=5$ system.

This solution is also known to interpolate between two well studied confining backgrounds, namely, the Euclidean one, studied by O. \'Alvarez, \cite{orlando}, and corresponding to the $r\rightarrow\infty$ limit, and the confining metric, studied by E. \'Alvarez and C. G\'omez, \cite{ag}, in the $r\rightarrow 0$ limit, which has been recently studied in detail in \cite{am} to show its confining character.

In this letter we will compute the four dimensional potential between two heavy sources $j\bar j$ when the effects of a dilaton field are taken into account. This will be shown to be associated to the one loop corrections of the potential in the sigma model approach. However, as will be explained later, there are other several possible sources of corrections. With this computation, we get some insight into the meaning of the fluctuations of the world sheet dictated from the beta functions.

The standard lore on the computation of the potential between heavy sources is based on interpreting the expectation value of the Wilson loop as a statistical average over random surfaces spanning the loop {\cite{wilson}}.

In the semiclassical limit, this average is represented through the minimal area surface spanned by the world sheet of the confining string \cite{loop, ry}. 

In the computation, an idealized rectangular static loop is posited in spacetime, by placing heavy sources at $x^1\equiv x=\pm L/2$, $x^i=0$ $(x^i\neq 1)$. The imbedding of the two-dimensional surface into the external space is defined by a unique function $r=\bar r(\sigma)$, determined by the Euler-Lagrange's equations, complemented with Dirichlet boundary conditions at the endpoints of the string, $\bar r(0)=\bar r(1)=\Lambda$.

The original computation of \cite{loop, ry} in AdS, started with a gauge group $U(N+1)$ broken to $U(N)\times U(1)$ by giving an expectation value to one of the scalars. This corresponds to having a $D3$-brane sitting at some radial position in AdS and at a point on $S^5$. The off-diagonal states, transforming in the {\bf N} of $U(N)$, get a mass proportional to the radial position. 

In order to get a non-dynamical source, we need to take the limit $m\rightarrow\infty$, which corresponds to work in the boundary of AdS. On the other hand, in this limit, the CFT is defined on an extension of Penrose's boundary, cf.\cite{ft,conjet}.

In this limit, the infinite mass of the sources causes a divergence in the potential and should be renormalized. This divergence comes form the fact that the string attached on the Higgsed $D3$-brane runs from the origin to the boundary at infinity. The configuration after the renormalization of the potential minimizes the action and resembles topologically half a cylinder. This is the configuration to be used in the computation of the Wilson loop.

Another way to deal with the singularity of the potential is by considering the Legendre transform of the area of the worldsheet. This is because these string coordinates obey the Neumann boundary conditions rather than the Dirichlet conditions \cite{ms}.

The inclusion of the dilaton field (\ref{dillor}) in the action takes into account only one of the possible contributions of ${\cal{O}}(\alpha')$: there are no corrections to the metric, and in this sense it can be considered as incomplete. 

However, the behaviour we find is very similar to the one expected in the Euclidean limit, first discovered in \cite{luscher} using a functional-integral approach in a semiclassical quantization of the Eguchi-Schild model, and by an argument based on an effective field theory governing the quantum dynamics at large distance scales of a thin flux tube linking the two sources. In four dimensions this is

\be
\label{lusch}
V(L)=\s L+\b-\frac{\pi}{12L},
\ee

\noindent where $\s$, the string tension, and $\b$ are model-dependent non-universal coefficients and the $1/L$ term is a long-distance effect.

This potential can also be obtained when the effect of quantum fluctuations of a Nambu-Goto like action is computed \cite{sonne}. On the other hand, it can be shown, \cite{arvi}, that the expression in (\ref{lusch}) is identical to the energy of the tachyonic mode of the bosonic string in flat spacetime with Dirichlet boundary conditions at $\pm L/2$. 

The existence of these corrections to the linear potential can be checked in lattice simulations. In fact, according to \cite{latt} there is some numerical evidence for a L\"uscher term associated with a bosonic string, as well as for glueball masses,  however these results are still not definitive.

\section{``Classical'' Wilson Loop}

The equivalence between working with the Nambu-Goto action in the static gauge or with the Polyakov action in the conformal gauge, i.e. the isothermal coordinate system, has been recently shown \cite{am}. The only difference between these two approaches is the existence of a second conserved charge when working in the conformal gauge, associated with the `00' component of the energy-momentum of the system, and a different value of the potential at infinity due to the boundary conditions.

We will work in the asymptotic limit $r\rightarrow 0$ in five dimensions, that is, in the regime of \cite{ag}, so the metric and the dilaton are

\be
\label{am}
ds^2=rdx^2+dr^2,
\ee
\be
\label{ad}
\Phi=-\log r.
\ee

In order to review the basic facts of such calculation, we will present the computation based on the Nambu-Goto action in the static gauge, that now, with the metric (\ref{am}) reads as

\begin{equation}
S=-\frac{T}{2\pi\alpha'}\int_{-L/2}^{L/2}\sqrt{r^2+r\left( {{\partial}_xr}\right)^2}dx.
\end{equation}

This action does not depend explicitely on $x$ and so its canonical conjugated momentum is a conserved quantity, namely

\begin{equation}
\frac{r^2}{\sqrt{r^2+r\left( {{\partial}_xr}\right)^2}}={\mbox{constant}}.
\end{equation}

As we have mentioned in the introduction, the standard configuration in this kind of computations is that of an open string with its ends attached to a four dimensional hyperplane, complemented with Dirichlet boundary conditions at $r=\Lambda$, so that the embeddind is half a topological cylinder. With this in mind, and denoting by $r_0$ the tip of this U-shaped string, we have

\begin{equation}
x={r_0}^{1/2}\int_{1}^{r/r_0}\frac{dy}{\sqrt{y\left( y^2-1\right)}}.
\end{equation}

Let us take firstly the limit $x \stackrel{r{\rightarrow}{\infty}}{\longrightarrow}\frac{L}{2}$. We can find the dependence of this quantity, interpreted as the energy, with the distance between the two sources $j\bar j$ as

\begin{equation}
\label{long}
\frac{L}{2}={r_0}^{1/2}\int_{1}^{\infty}\frac{dy}{\sqrt{y\left( y^2-1\right)}},
\end{equation}

\noindent or equivalently

\begin{equation}
r_0\propto L^2.
\end{equation}

Now we can compute the potential between $j\bar j$ as

\begin{equation}
\label{ener}
E=\frac{S}{T}=\int_{-L/2}^{L/2}dx\frac{r^2}{r_0},
\end{equation}

\noindent this is

\begin{equation}
\label{potnodil}
E={r_0}^{3/2}\int_{1}^{\infty}\frac{y^2dy}{\sqrt{y\left( y^2-1\right)}}.
\end{equation}

In this way we finally obtain the dependence on $L$ of the energy between the heavy sources as

\begin{equation}
E\propto L^3.
\end{equation}

This computation has been made by imposing boundary conditions on the hyperplane at infinity. However, one can obtain more information by introducing a cutoff at a value $r=\Lambda$. In this case, equation (\ref{long}) also tells us that for each $L$ there are two available configurations, thus dividing the $r$ coordinate in two regions, whose asymptotic regimes correspond to the limits $\Lambda\rightarrow\infty$, {\sl region I}, and $\Lambda\rightarrow r_0$, {\sl region II} (\cite{am, ag2, kogan}).

It has also been shown that region II corresponds to a linear confining branch, where the potential behaves approximately as

\be
\label{potenconf}
V=\frac{L\Lambda}{3\pi l_s^2},
\ee

\noindent and moreover, that this is the energetically favoured branch for all the range of the coordinate $r$, but near the naked singularity at the origin,  where this configuration dissapears and we are only left with the configuration of region I.

\begin{figure}[h]
\begin{center}
\leavevmode
\epsfxsize=12cm
\epsffile{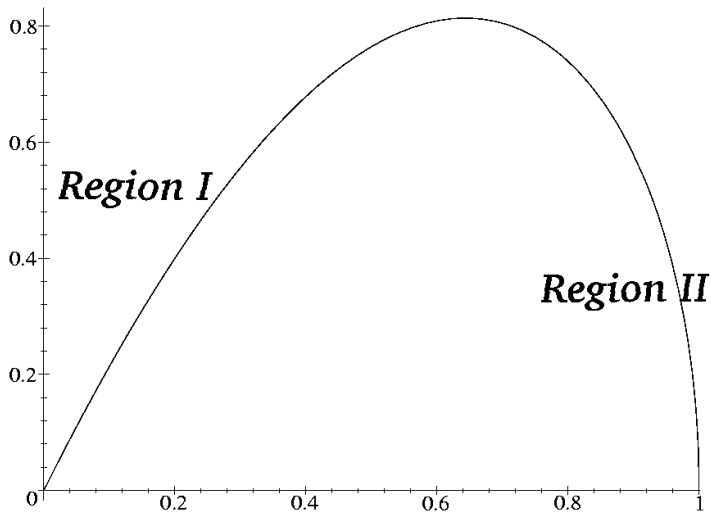}
\vspace{-1cm}
\caption{\it $\frac{L}{\sqrt{\Lambda}}$ vs. $\sqrt{\frac{\phi_0}{\Lambda}}$}
\label{f1}
\end{center}
\end{figure}

We will explore the role played by the singularity at the end of the next section. Let us only mention that the existence of a maximun in the value of $L$, may indicate the existence of an effective horizon covering the singularity \cite{ag2, s}, in analogy with the Schwarzchild-anti de Sitter case studied in \cite{ry}.

\section{Quantum fluctuations}

In this section we will include the dilaton field to compute the potential between the two heavy sources. As explained in the introduction, this would correspond to a first order perturbation in $\alpha'$, and could be considered as the minimal correction to the area law behaviour {\sl ansatz}, mainly due to the fact that the value of this field is taken from the beta equations at the same order than the solution (\ref{am}) for the metric studied in the previous section.

In this sense, this approach is different of those considered in the literature. For example, in the AdS case studied in \cite{go}, the minimal surface ansazt was employed and then the saddle point field for large interquarks distances was taken in order to compute the potential. In this case the conclusion was that the effective flatness of AdS space, in the strong-coupling limit, suggested that the fermi and ghost degrees of freedom contributed to the L\"uscher term as in flat space. However, despite the right magnitude of the factor, its sign was the opposite.

There are other contributions of the same order in $\a'$, for example, perturbations of the metric as in \cite{sonn} or sigma model solutions to the beta equations to the next to leading order in $\a'$ as in \cite{ahd}.

As shown in \cite{ahd}, there are two equivalent points of view in order to consider the different $\a'$ corrections. Starting from the background given by (\ref{am}), (\ref{ad}), keeping the dilaton fixed, the perturbed metric will be

\be
\label{pm}
ds^2=\left(r+\left(\frac{l_s^2}{l_c^2}\right)^2\left(c_1r-\frac{1}{32r}\right)\right)\eta_{\m\n}dx^\m dx^\n+l_c^2\left(1+\left(\frac{l_s^2}{l_c^2}\right)^2\left(c_2+\frac{1}{2r^2}\right)\right)dr^2,
\ee

\noindent where $c_1$ and $c_2$ are two integration constants and $l_c$ is a lenght scale whose role will be explained at the end of this section. This is equivalent, up to a change of coordinates, to modify the term conformal to the Minkowski(Euclidean) metric in the background, keeping the holographic term fixed. In this way, the dilaton gets shifted to

\be
\label{dm}
\Phi=-\log(r)+\frac{l_s^2}{l_c^2}\left(\frac{c_2}{2}+\frac{c_3}{2r}-\frac{1}{4r^2}\right),
\ee

\noindent where $c_3$ is an arbitrary constant and $c_2$ indicates the freedom to choose a zero point on the original dilaton, due to the fact that it only appears in the beta equations in the form of a derivative.

We may argue that this equivalence is exact to all orders in the sigma model expansion, altough it has only been checked at this order. In this manner, taking the first point of view, the solution (\ref{ad}) for the dilaton may be considered as an exact solution.

Moreover, the corrections to the metric vanish in the limit $r\rightarrow\infty$, so in the IR, the most important correction comes from the dilaton. In the UV limit this may not be the case. In the UV the curvature grows and the other terms in the sigma model expansion become important. In fact, as we approach the singularity, all the terms in the expansion should be included.

As we will see, in this UV limit appear certain singularities in the potential. These singularities are not too reliable, and its origin would be precisely what we have just commented on the truncation of the $\a'$ expansion.

We will work with the Polyakov action, that is

\begin{equation}
S=\frac{1}{4\pi\alpha'}\int_{M}d^2{\sigma}g^{1/2}\left[ g^{ab}G_{{\mu}{\nu}}(X){\partial}_aX^{\mu}{\partial}_bX^{\nu}+{\alpha}'R^{(2)}{\Phi}(X) \right],
\end{equation}

\noindent where $R^{(2)}$ is the world sheet Ricci scalar, which we take as defined by the induced metric over the worldsheet, and the dilaton is taken as in (\ref{ad})

\begin{equation}
\Phi(X)=-\log r,
\end{equation}

\noindent therefore we can write the action as

\begin{eqnarray}
\label{accdila}
S=\frac{T}{4\pi\alpha'}\int_{-L/2}^{L/2}dx\Biggr{\{} 2-\alpha'\log r\frac{2r^2{\partial}^2_xr-2r({\partial}_xr)^2-({\partial}_xr)^4}{2r^2\left[ r+ \left( {\partial}_xr \right)^2 \right]^2} \Biggr{\}}\sqrt{r^2+r{\left( {\partial}_xr \right)}^2},
\end{eqnarray}

\noindent where we have made use of the static gauge, that is, $\sigma=x$ and $\tau=t$. This can be done because the we are taking the metric as a non-dynamical variable, so we are simply writing the Nambu-Goto action in another way{\footnote{I thank the referee for pointing this fact}}. 

Now, as in the previous case, the canonical momentum conjugated to $x$ is a conserved charge. To determine this constant we proceed as in the previous section, i.e. calling $r_0$ as the minimum of $r$, we have

\be
\label{const}
\frac{2r^3+2r^2(\partial_xr)^2-\alpha'(\partial_xr)^2}{r^{1/2}(r+(\partial_xr)^2)^{3/2}}=2r_0.
\ee

An interesting feature of this equation is that the presence of the dilaton does not affect the value of $r_0$, i.e. it has no corrections of order $\alpha'$. This is easily seen in the solution of this equation, that can be written as

\be
x=\pm r_0^{1/2}\int_1^{\Lambda/r_0}\frac{dy}{\sqrt{y(y^2-1)}}.
\ee

When solving to get this equation, we have neglected terms ${\cal{O}}(\alpha'^2)$. This integral can be easily computed in terms of elliptic functions as

\be
\label{longi}
L=\pm 2r_0^{1/2}\frac{1}{\sqrt{2}}F\left( \cos^{-1}\sqrt{\frac{r_0}{\Lambda}},\frac{1}{\sqrt{2}}\right).
\ee

So even if we add this dilaton field, the discussion of the previous section concerning the two different regions determined by the value of $\Lambda$ is still valid. This can be seen in Fig.(\ref{f1})

Now we go on to compute the potential between the heavy sources placed at $r=\Lambda$. This will be done following the definition

\be
V(L)=\lim_{T\rightarrow\infty}\frac{S}{T},
\ee

\noindent but now we have to be careful with the interpretation one is to give to this prescription. With the dilaton term, the on-shell solutions of the action (\ref{accdila}) cannot probably be interpreted as the minimal area surfaces. The potential takes the form

\be
\label{vl}
V(L)=\frac{2r_0^{3/2}}{\pi\alpha'}\int_1^\aleph\frac{z^4dz}{(z^4-1)^{1/2}}+\frac{\log(r_0)}{2\pi r_0^{1/2}}\int_1^\aleph\frac{z^2(z^4-3)dz}{(z^4-1)^2}+\frac{1}{\pi r_0^{1/2}}\int_1^\aleph\frac{\log zz^2(z^4-3)dz}{(z^4-1)^2},
\ee

\noindent where $\aleph=\sqrt{\Lambda/r_0}$. When evaluating this integral, one must be careful, because it has an infrared divergence, so we will integrate from $1+\epsilon$ to $\aleph$.

This divergence may be associated with the absence of corrections to the loop in this integration limit as seen in (\ref{longi}), i.e. to the fact that the value of $r_0$ remains unchanged in presence of the dilaton field. Later we will focus on the possible cancellation of these divergences.

The first integral in the r.h.s. of (\ref{vl}) is exactly the contribution to the $j\bar j$ potential studied in the previous section, eq.(\ref{potnodil}), so let us concentrate on the extra terms due to the dilaton

\be
V_{\Phi}=\frac{\log(r_0)}{2\pi r_0^{1/2}}\int_{1+\epsilon}^\aleph\frac{z^2(z^4-3)dz}{(z^4-1)^2}+\frac{1}{\pi r_0^{1/2}}\int_{1+\epsilon}^\aleph\frac{\log zz^2(z^4-3)dz}{(z^4-1)^2}.
\ee

These integrals can be computed as

\bea
V_{\Phi}=\frac{\log(r_0)}{2\pi r_0^{1/2}}\left[\frac{x^3}{2(x^4-1)}+\frac{3}{8}\log\left(\frac{x-1}{x+1}\right)\right]_{1+\epsilon}^\aleph+\frac{1}{\pi r_0^{1/2}}\Biggr[\frac{1}{8}\log\left(\frac{x+1}{x-1}\right)+\frac{x^3\log x}{2(x^4-1)}+\hspace{0.5cm}\\
\nonumber \\
\nonumber \frac{3\pi}{16}\log(1+x^2)-\frac{3}{8}\log(x)\log(1+x)-\frac{1}{4}{\mbox{arctan}}(x)+\frac{3}{8}\left(\int_x^{x+1}\frac{\log t}{1-t}dt-i\int_{-ix}^{ix}\frac{\log t}{1-t}dt\right)\Biggr]_{1+\epsilon}^\aleph.
\eea

As in the previous section, let us first evaluate this equation in the limit $\Lambda\rightarrow\infty$. In this case we find

\be
\label{enerpot}
V_{\Phi}=\frac{\log(r_0)}{2\pi r_0^{1/2}}\left[\frac{3\pi-6+6\log(2)}{16}\right]+\frac{1}{\pi r_0^{1/2}}\left[\frac{3\pi^2-2\pi-2+24{\cal{C}}-6\log(2)(\pi^2+\pi+2)}{32}\right],
\ee

\noindent where ${\cal{C}}$ is the Catalan number and we have neglected the terms ${\cal{O}}(\log(\epsilon))$. The dependence of the potential with $L$ is included in the value of $r_0$, which is read from (\ref{longi}). In this limit this is simply

\be
r_0=\frac{L^2}{2K^2(1/\sqrt{2})}.
\ee

Inserting this in (\ref{enerpot}) we have the potential

\be
\label{finpot}
V_{\Phi}\approx\frac{2\sqrt{2}(\log(L)-1)}{\pi L}-\frac{3}{5\pi L}+({\cal{O}}\log(\epsilon)).
\ee

Let us now go on to examine the region II, i.e. the limit $\Lambda\rightarrow r_0$. We must perform an integral from $1+\epsilon$ to $1+\delta$, due to the infrared divergence of the potential. Forgetting these effects, we can write down the potential as

\be
V_{\Phi}=-\frac{\log(\Lambda)}{16\pi\sqrt{\Lambda}}-\frac{\d^{1/2}}{\pi L}\left[\frac{\sqrt{2}}{16}\log(\Lambda)+\sqrt{2}\log(L)+\frac{6}{5}\right].
\ee

This branch has been shown to correspond to an energetically favoured linear confining regime (\ref{potenconf}), so we can write down the potential between the heavy sources with a leading order correction due to fluctuations of the string given by the presence of the dilaton as

\be
\label{pdef}
V\approx\frac{L\Lambda}{3\pi l_s^2}-\frac{\log(\Lambda)}{16\pi\sqrt{\Lambda}}-\frac{\d^{1/2}}{\pi L}\left[\frac{\sqrt{2}}{16}\log(\Lambda)+\sqrt{2}\log(L)+\frac{6}{5}\right],
\ee

\noindent plus corrections ${\cal{O}}(\log(\epsilon))$ and ${\cal{O}}(\log(\delta))$. 

An interesting feature of this potential is that it shows that the string tension has not been changed by the presence of the dilaton. 

On the other hand, there is a dependence in the cutoffs introduced in the computation, that is, in $\Lambda$ and $\d$, which may correspond to choosing a particular value of the coupling constant.

\begin{figure}[h]
\begin{center}
\leavevmode
\epsfxsize=7cm
\epsffile{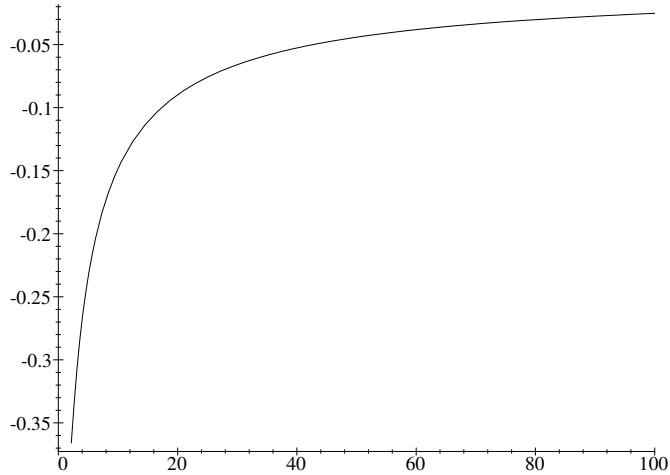}
\vspace{-1.5cm}
\caption{\it $V_{\Phi}$ vs. $L$ at $\Lambda=10$}
\label{f2}
\end{center}
\end{figure}

In Fig.(\ref{f2}) we have plotted the $L$ dependent term of $V_\Phi$, where we can see that the dependence with L is of the kind expected.

The corrections ${\cal{O}}(\log(\delta))$ in (\ref{pdef}) are not important due to the fact that the limit $\delta=0$ will only be taken at the origin, $r=0$, where the naked singularity lives, so we will never reach it. This is only at the origin we can have $r_0=\L$, which corresponds to $L=0$, as can be seen in Fig.(\ref{f1}).

This is the reason why Fig.(\ref{f2}) can only be but for values of $L/l_c$, where $l_c$ is a length scale, not too small. For example, when $\Lambda=10$, $L$ cannot be smaller than order 0.2 in $l_c$ units. In this ratio between the length of the loop and the length scale $l_c$, this last is given by

\be
l_c=\frac{1}{r_0}\sqrt{\frac{6\a'}{21}},
\ee

\noindent and has been set to unity in this letter. 

This same phenomenon occurs in absence of the dilaton, as stated in \cite{am}, and could be thought of as if in this limit the theory became free.

As we move towards the origin, the length between the heavy sources decreases. As seen on \cite{ahd}, there is a limit in the holographic coordinate, namely $r>l_c/\sqrt{\a'}$, and beyond it, this level of approximation in the sigma model is nonsense. This scale would correspond to $L/l_c\lesssim 0.2$. Moreover, we can find a minimum of the potential that, in the case $\Lambda=10$, is

\be
V'_\Phi=\frac{\sqrt{2}/16\log(10)+\sqrt{2}\log(L)+6/5}{\pi L^2}-\frac{\sqrt{2}}{\pi L^2}.
\ee

This tells us that there is an equilibrium point for this part of the potential at

\be
\frac{L}{l_c}\approx 1
\ee

\noindent and this equilibrium point corresponds to a free theory. The linear dependent part of the potential produces just a shift in this value.

\section{Concluding remarks}

In this letter we have studied the leading corrections to the static gauge potential between heavy sources as dictated by the beta equations. In this sense we have computed a limit which departs somewhat from the decoupling limit defined as the limit $\a'\rightarrow 0$.

The case studied is one of the asymptotic limits of the background introduced in \cite{agh}, and the main interest of the computation lies on the prediction of a L\"uscher like term in the potential. 

To compute the potential between heavy sources a spatial Wilson loop has been considered, in such a way that the Wilson loop operator exhibits some kind of area law, i.e.

\be
<W(C)>\sim\exp(-TV(L))\sim\exp(-\sigma A(C)),
\ee

\noindent where $A(C)$ is the area enclosed by the loop. 

In this letter the dilaton field has been introduced, so the action in principle cannot be interpreted as the area of the world sheet. However, we can think of this as a small perturbation to the area of the worldsheet and, in this sense, follow the same prescription.

The solution coming from the dilaton can be considered as an exact solution, while the other corrections in $\a'$ can be absorbed in the form of corrections to the metric. Moreover, it has been shown that this correction is dominant in the IR, while in the UV limit the analysis is not conclusive, due to the presence of the naked singularity, which forces us to include more terms of the sigma model expansion.

This ambiguity translates in the presence of singularities in the potential, that may be cancelled when all the other terms will be included.

The potential obtained resembles the one obtained by L\"uscher-Symanzik-Weisz but now the coefficient of the $1/L$ term is non-universal, but it has instead a dependence on both $\Lambda$, the scale at which we placed the loop, and $\delta$, a cutoff introduced in the computation. This is a dependence on the coupling constant and may indicate a screening of the charges in the IR limit.

It has also been found a stable equilibrium point in the potential coming from the dilaton in the vicinity of the origin, that is, near the naked singularity. In this way it is temptative to think about the singularity as a point where the dual gauge theory becomes free.

\section*{Acknowledgements}

I would like to thank E.~\'Alvarez, C.~G\'omez, L.~Hern\'andez, D.G.~Cerde\~no and N.~Alonso for useful discussions and for the critical lecture of the manuscript. I would also like to thank J.~Delclaux for colaborating in the first stages of this letter and for very helpful discussions.


\end{document}